\documentclass[aps]{revtex4}

\usepackage{amssymb}
\usepackage{graphicx}
\usepackage{subfigure}

\newif\ifpdf
\ifx\pdfoutput\undefined
\pdffalse
\else
\pdfoutput=1 
\pdftrue
\fi
\ifpdf
\else
\fi

\begin{document}

\draft
\title{Stromatolites: why do we care?}
\author{Massimiliano Ignaccolo$^{1,}$\footnote{Corrispondent Author. Email mi0009@unt.edu (active till the end of September) \textbf{or} m\_ignaccolo@yahoo.it}}
\author{Arne Schwettmann$^{1}$}
\author {Roberto Failla$^{1}$}
\author{Michael C. Storrie-Lombardi$^{2}$}
\author{Paolo Grigolini$^{1,3,4}$}
%\author{Guenter Gross$^{4}$}
\affiliation{$^{1}$Center for Nonlinear Science, University of North Texas,\\P.O. Box 311427, Denton, Texas 76203-1427, USA}
\affiliation{$^{2}$Kinohi Institute, Pasadena, CA 91101 USA}
\affiliation{$^{3}$Dipartimento di Fisica dell'Universit\`{a} di Pisa and INFM, Via Buonarroti 2, 56127 Pisa, Italy}
\affiliation{$^{4}$Istituto dei Processi Chimico Fisici del
CNR,Area della Ricerca di Pisa, Via G. Moruzzi 1, 56124 Pisa,
Italy}
%\address{$^{4}$ University of North Texas Center for NetworkNeuroscience}
\date{\today}
%\maketitle
%%%\begin{document}
\begin{abstract}
We apply the method of Diffusion Entropy (DE) to the study of
stromatolites by means of a two-dimensional procedure that makes it
possible for us to compare the DE analysis to the results of a
compression method. As done with the compression method, we analyze
two pairs of samples, one biotic and the other a-biotic. Each pair
consists of a target, the putative stromatolite sample, and of its
surrounding matrix. We use two different procedures, referring to
single colors and to a color combination, respectively. We apply the
DE method to both procedures and we find the same result, this being
that the scaling index of the time series stemming from the biotic
target yields a scaling index larger than the scaling indices of the
other three curves. We argue that the DE analysis confirms the results
of the compression method.
\end{abstract}
\maketitle

\section{Introduction}\label{intro}
Let us explain the meaning of the title of this paper. One of the
papers of these Proceedings \cite{michael} is devoted to the
intriguing problem of distinguishing biotic from a-biotic
stromatolites. Why do we care about the conclusion of
Ref. \cite{michael}? This is the meaning of the question posed by the
title of this paper. To answer this question, we have to draw the
reader's attention to an earlier work \cite{giulia}, devoted to the
problem of assessing the complexity of a process, as it is mirrored by
a time series derived in some way from the process under study. The
title of the paper of Ref. \cite{giulia}: " Compression and diffusion:
a joint approach to detect complexity" is self-explanatory. In fact,
the authors of Ref. \cite{giulia} discussed two methods of analysis of
time series, the former with a mathematical foundation, and the latter
with a physical (dynamic) foundation. The former method rests on the
concept of computable information content and is, consequently,
closely related to the compression algorithm used in
Ref. \cite{michael}. The second method rests on converting first the
time series to study into a diffusion process. Then, as a second step,
we evaluate the entropy of this diffusion process \cite{giulia}, this
being the reason why the method is called Diffusion Entropy (DE)
method. We refer the reader to the paper of Ref. \cite{giulia} for
details on both method.  Here we limit ourselves to observing that the
connection between the two methods of analysis is explained as
follows.  If a time series is found to be compressible, this is an
indication that there is strong correlation. This strong correlation
means that the diffusion process generated by the first step of the DE
method must reach an anomalous scaling regime, with a scaling
parameter $\delta$ that is expected to be different from the scaling
of diffusion process that would be generated by a random time
series. We expect that the more compressible the time series is, the
larger is the deviation of the diffusion scaling from the ordinary
value ($\delta = 0.5$).  This is the main reason why we care about the
results of paper
\cite{michael}, which uses the compression method to study
stromatolites. If the arguments of Ref. \cite{giulia} are correct, and
of general validity, then the DE method applied to the same
stromatolites as those studied in Ref. \cite{michael}, should yield for
the biotic stromatolite a scaling parameter larger than that stemming
from the a-biotic stromatolite.  However, to get this wished result,
which would support the conclusion of the earlier paper \cite{michael}
and, at the same time, would confirm that the arguments of
Ref. \cite{giulia} have a general validity, we have to adapt the DE
method to the specific requests posed by the analysis of digital
images of polished stromatolite slabs. We solve this problem in
Section 2. Section 3 is devoted to illustrating the results and
Section 4 to the concluding remarks.

\section{Methods}\label{methods}
\begin{figure}[h]
\includegraphics[angle=-90,width=2in] {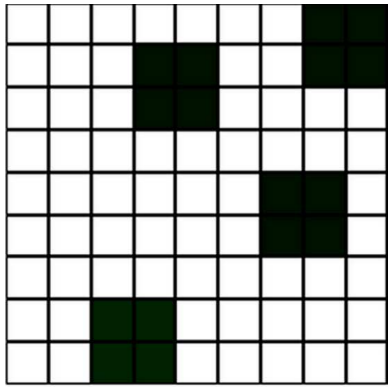}
\caption{Example of the procedure for extending the DE method
to the analysis of data represented by matrixes. The small black
squares represent some of the many ways of locating a square of size
$2$ inside a matrix of dimension $9 \times
9$.}
\label{figure1}
\end{figure}

In this Section we describe, briefly,
how to adapt the DE method to the analysis of digital RGB uncompressed
TIFF format images of polished stromatolite slabs. For technical
details on the meaning of the acronyms, we refer the reader to the
paper of Ref. \cite{michael} and to the papers there quoted. Each
pixel of a digitalized RGB image is labeled with three integer
numbers, representing the intensity of each of the three fundamental
colors: a value of $0$ indicates the absence of the color, while a
value of $255$ indicates the presence of the color with the maximum
possible intensity, according to the RGB scale. Therefore each RGB
image is represented by three $M \times N$ different matrixes, with
$M$ and $N$ denoting the dimensions, in pixels, of the image.  The DE
method for the analysis of time series, namely an ordered sequence of
data represented by an one dimensional array, rests on using the time
series itself, to create a diffusion process and to monitor the
entropy increase as a function of time. To accomplish this, we move a
window of increasing size $l$ through the time series, summing all the
data therein contained. If $N$ is the length of the time series and
$l$ is the size of the window adopted, $N-l+1$ different data are
obtained through the moving and summing procedure. This numbers are
considered as the possible positions that an imaginary walker can
assume after moving for a time $l$, and used to define the probability
density function (pdf) of the diffusion process at this time. Finally,
we evaluate the Shannon entropy of the pdf. The generalization of this
procedure to the case where the data to examine are represented by a
matrix is simple. The window of size $l$ to move through the sequence
of data, is substituted by a square of size $l$ to be moved through
the matrix. Fig. \ref{figure1} illustrates how to apply this procedure
to a square of size $l=2$ and, thus, to a matrix $9 \times 9$. The sum
of all the elements inside a square has to be considered as one of the
possible positions that an imaginary walker can assume after a time
$l^{2}$. With this modification it is possible to use the data
contained in a matrix to build up a one-dimensional diffusion
process. A digital RGB image stems from three different matrixes and
we are naturally led to two possible ways of applying the DE method to
the image. The first one is to create a one-dimensional diffusion
process for each one of the three matrixes representing the intensity
of the three fundamental colors, and to examine the dependence of the
diffusion entropy on the color. We shall refer to this analysis as the
``single color DE image analysis''. The second one is to use the three
matrixes to create a three-dimensional diffusion process and to
evaluate the change in time of its entropy. This is done using the
three matrixes of an RGB image file, to determine the movement of the
imaginary walker in the $(x, y, z)$ space, with the x, the y and the
z-axis, referring to the red, green and blue color, respectively. This
last technique will be referred to by us as the ``global DE image
analysis''.

%%%%%%%%%%%% Figure 1

\section{Results}\label{results}

\begin{figure}[h]
\includegraphics[angle=-90,width=3.5in] {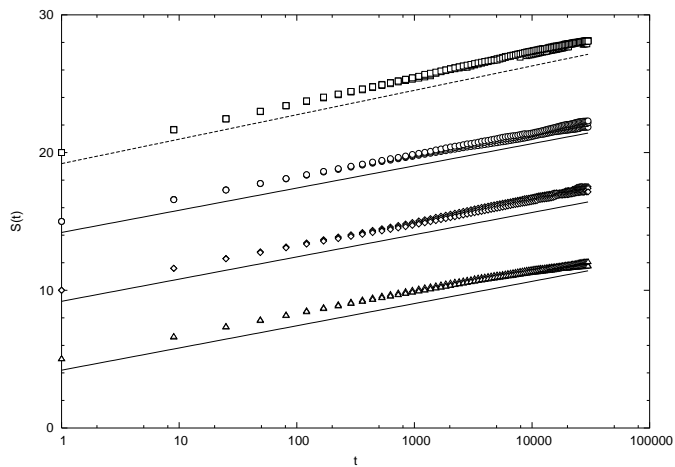}
\caption{Results of the single color DE
image analysis. The entropy $S(t)$ as a function of time $t$ in a
logarithmic scale. For each image the results relative to the red, green and blue component of the image are plotted. The square indicates the results  relative to the biotic target, the circles those relative to the matrix surrounding the biotic target, the diamonds those relative to the a-biotic tartget and the triangles those of the matrix surrounding the a-biotic target. Finally, the full and the dashed lines indicate a straight line, in logarithmic time scale, of slope $0.7$ and $0.77$, respectively.} 
\label{figure2}
\end{figure}

In order to establish, hopefully, the biogenicity of a given
stromatolite through statistical analysis of the corresponding RGB
images, the authors of Ref. \cite{michael} decided to compare the
results of the compressibility of putative biotic and a-biotic
stromatolites with the compressibility of the rock matrix surrounding
the samples. Here, we shall follow the same strategy for the results
of the single-color DE image analysis and global DE image analysis. We
explore the same database of Ref. \cite{michael} and, for brevity, we
report the results relative to just four images, a putative biotic
target, the surrounding matrix of the biotic target, a putative
a-biotic target and its surrounding matrix. Fig. \ref{figure2} shows
the results of the application of the single-color DE image
analysis. For clarity, the curves relative to different images are
shifted to different initial values. Moreover, another shifting
procedure is applied to the three curves relative to the three
different color matrixes of a single image, so that only the rates of
increase of diffusion entropy are compared. We see that the three
different colors, for all the images, produce, approximately, the same
curve and that only the curves of the putative biotic image yield a
quantitatively different rate of entropy increase, as indicated by the
corresponding slopes, which are different from the others.

\begin{figure}[h]
\includegraphics[angle=-90,width=3.5in] {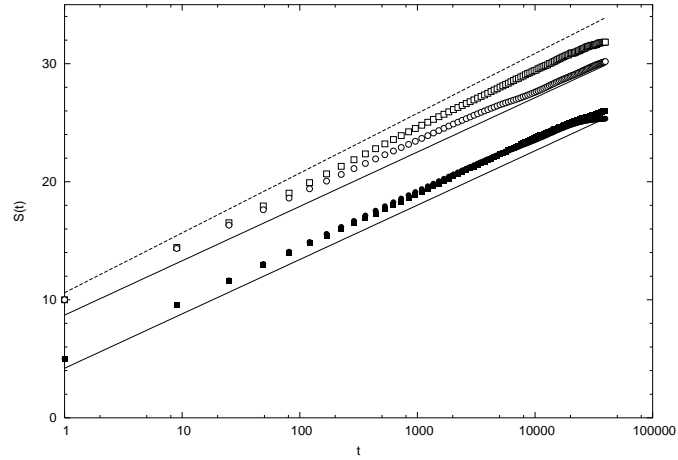}
\caption {Results of the global DE image
analysis. The entropy $S(t)$ as a function of time $t$ in a
logarithmic scale. The empty squares and circles indicate the results relative to the biotic target and its surrounding matrix, while the black ones the results relative to the a-biotic target and its surrounding. Finally the full line and dashed lines indicate a straight line, in logarithmic time scale, of slope $2$ and $2.2$, respectively.}
\label{figure3}
\end{figure}

Fig. \ref{figure3} shows the results of the global DE image
analysis. Again, a procedure of shifting of the curves is adopted, so
as to compare among themselves only the rates of entropy increase, in
the linear-log representation. We see clearly that the diffusion
process generated by the putative a-biotic sample and the one
generated by its surrounding matrix yield, virtually, the same rate of
entropy increase. It is evident, though, that the putative biotic sample
yields a slope distinctly larger than that stemming from its
surrounding matrix, a slope which, is, on the contrary, close to that
of both a-biotic target and a-biotic surrounding. Fig. 3 shows that
the biotic target has a slope close to $\delta = 2.2$, while the other
three curves yield a slope close to $\delta = 2$. Note that an
entropy increase with a slope $\delta = 1.5$ is the result expected
for an ordinary three-dimensional diffusion.

\section{Concluding Remarks}

We see that the extension of the DE method to the analysis of
two-dimensional images yields results supporting the conclusion of the
paper of Ref.\cite{michael}.  The most compressible two-dimensional
image yields the largest scaling. As to the relevance of this
conclusion for the challenge posed by the search for a way to
distinguish biotic from a-biotic stromatolites, we limit ourselves to
the following remark. We see that, although the biotic target yields
the highest scaling, also the scaling of both biotic and a-biotic
surroundings are larger than the ordinary scaling. This means that
complexity is a general property. It is not only a property of living
systems, it is rather a property of the so called \emph{living state
of matter} \cite{marco,gerardo}. Thus, the fact that distinguishing
biotic from a-biotic stromatolites is a challenging issue, is a way to
answer a provocative question raised by the author of
Ref. \cite{gunter}.  Pete Gunter \cite{gunter} writes: " How odd it is
that Ð on this planet at least Ð life emerged at virtually the same
time as rocks." We think that the field of Complexity is affording an
answer to this question. Life and rocks are perceived as foreign the
one to the others, if the vision of ordinary statistical mechanics is
adopted. Ordinary statistical mechanics is not a fair representation
of reality, but only a condition mirroring one of the two basic
properties of nature: randomness, the other being order. The balance
of randomness with order yields a condition intermediate between the
dynamic and the thermodynamic state \cite{gerardo}, with phenomena
such as aging that according to the traditional wisdom seem to be
peculiar to living systems. On the contrary, as proven in
Ref. \cite{gerardo}, these phenomena are manifested also by non-living
systems. So, a new perspective, moving from Bergson \cite{gunter} ,
benefiting from the pioneering work of Prigogine \cite{gunter,marco},
and yielding, finally, the concept of Living State of Matter
\cite{marco, gerardo}, can also be used to explain why the
stromatolite problem is difficult, and so challenging, and much more
research work is required to establish how general are the conclusion
of this paper and of the companion paper of Ref. \cite{michael}.
\newline 
\newline
PG and MI acknowledge support from ARO, through Grant DAAD19-02-0037.

%%%%%%%%%%%% Figure2

%%%%%%%%%%%%%%%%%% Figure 3

\end{document}